# Non-autoregressive real-time Accent Conversion model with voice cloning

*Vladimir Nechaev, Sergey Kosyakov*

Ivanovo State Power Engineering University
nechaev@gapps.ispu.ru

**ABSTRACT**

Currently, the development of Foreign Accent Conversion (FAC) models utilizes deep neural network architectures, as well as ensembles of neural networks for speech recognition and speech generation. The use of these models is limited by architectural features, which does not allow flexible changes in the timbre of the generated speech and requires the accumulation of context, leading to increased delays in generation and makes these systems unsuitable for use in real-time multi-user communication scenarios. We have developed the non-autoregressive model for real-time accent conversion with voice cloning. The model generates native-sounding L1 speech with minimal latency based on input L2 accented speech. The model consists of interconnected modules for extracting accent, gender, and speaker embeddings, converting speech, generating spectrograms, and decoding the resulting spectrogram into an audio signal. The model has the ability to save, clone and change the timbre, gender and accent of the speaker's voice in real time. The results of the objective assessment show that the model improves speech quality, leading to enhanced recognition performance in existing ASR systems. The results of subjective tests show that the proposed accent and gender encoder improves the generation quality. The developed model demonstrates high-quality low-latency accent conversion, voice cloning, and speech enhancement capabilities, making it suitable for real-time multi-user communication scenarios.

## 1. INTRODUCTION

An accent is an essential feature of speech. It can be native, influenced by various regional and cultural factors, or foreign [1]. A foreign accent differs from a native one at both the segmental (phonemes) and suprasegmental (intonation, stress, rhythm) levels [2]. A foreign accent appears when a speaker of one language (L1) speaks another, non-native or second language (L2) [3]. L2 speech may be less clear to native speakers compared to equivalent L1 speech [4], leading to reduced understanding and trust, negative attitudes towards the speaker, and potential discrimination [5, 6, 7].

Native speakers perceive grammar and fluency on one hand, and accent on the other, as separate entities. Improving one of them enhances the overall perception of L2 speech by native speakers [4, 8]. However, achieving perfect L1 pronunciation by a non-native speaker is practically difficult due to differences in phonetic interference and perception by native speakers [9]. Therefore, automated conversion of L2 speech to L1 speech is an important task to improve communication quality between people [10].

Accent conversion systems can be used in language learning [10, 11], for re-recording and improving the quality of previously recorded speech [12], to enhance the performance of existing speech recognition systems [13], and in call centers serving foreign clients [14].

When converting accents, it is important to preserve the speaker's unique vocal characteristics (timbre, pitch, volume), which means voice cloning is necessary [15]. This involves modifying the segmental and suprasegmental features related to the foreign accent [16, 17, 18]. In real-time scenarios, such as voice communication using high-speed connections, it is crucial to ensure minimal delays in generating and transmitting the processed speech [19, 20, 21].

### 1.1. Related work

Early methods of accent conversion at the generation stage require reference L1 examples that match the L2 speech [16, 22, 23, 24]. This means that each L2 phrase needs a corresponding L1 phrase. The practical application of such models is limited due to the lack of data to cover all possible speech variations. These approaches require significant resources for data collection and processing, which increases the time and cost of developing such systems. Additionally, relying on strict paired examples can reduce the versatility and scalability of the technology, limiting its ability to adapt to new accents or speech styles that were not included in the original dataset.

Subsequent developments have overcome this limitation, and reference examples are no longer required during inference [17, 18, 25, 26, 27, 28]. However, models [17, 26] still require parallel datasets containing matching L2 and L1 phrases during the training phase. Obtaining enough of such data is difficult and expensive. Methods [17, 26] use autoregressive recurrent neural networks, complicating the training process. Method [25] uses pre-trained neural networks for text-to-speech conversion, requiring a separate model for each target speaker to maintain unique vocal characteristics, making multi-user usage challenging. Methods [27, 28] predict the duration of each generated phoneme, altering the original L2 speech duration and speaker identity, and require context accumulation, increasing generation time and complicating real-time use. Method [18] lacks these drawbacks but is limited to a fixed set of accents, with the accent identifier needing to be determined during model training. This complicates data preparation and the application of the model to previously unincluded accents.

This work presents an accent conversion method that addresses all the aforementioned issues and limitations. It enables the transformation of any speech from L2 to L1 without the need for reference examples and parallel data during training and generation stages, significantly simplifying, reducing the cost, and speeding up the system's adaptation to new accents. By introducing an accent and gender encoder based on voice, it overcomes the limitation of a fixed set of accents in method [18]. The model extracts voice characteristics related to L1 and L2 accents and gender, representing them as fixed-length vectors, which are then used in the speech-to-phonetic and spectrogram conversion stages.

A demonstration of the proposed method is available on YouTube[1].

The architecture of the developed model utilizes neural network layers of the following types: feedforward [29], convolutional [30], and a Transformer encoder with an attention mechanism [31]. The proposed model is non-autoregressive and does not use recurrent networks in its architecture [32], which significantly speeds up the training process, allows real-time accent conversion, and avoids the error accumulation effect associated with sequential output generation.

The proposed method also includes a voice cloning algorithm using a speaker identification model, which converts individual vocal characteristics into a fixed-length vector. This vector, along with vectors for accent, gender, and fundamental frequency [33], is used in the transformation stage from the phoneme matrix to the spectrogram. This preserves the speaker's identity even after accent conversion. This is especially important in situations where it is necessary to retain the emotional tone, expressiveness, and unique speech characteristics. Additionally, the method allows real-time modification of voice characteristics, such as accent, timbre, and gender-related features, by copying the corresponding characteristics from an audio sample. This makes it applicable to a broader range of scenarios than previous methods.

---
[1] https://youtu.be/5obYWIugKS0

## 2. MODEL DESCRIPTION

The proposed accent conversion method comprises several interconnected modules unified into a single end-to-end architecture. These models are used for accent and gender identification, speaker recognition, speech-to-phonetic token conversion, spectrogram generation, and decoding the resulting spectrogram into an audio signal. Figure 1 shows the overall interaction scheme of these models during the inference stage, which is the generation of the output L1 audio.

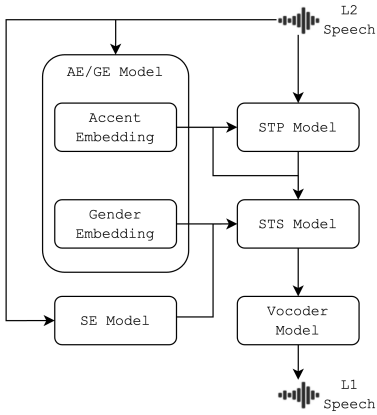

**Fig. 1.** General scheme of accent conversion model inference with voice cloning.

The incoming audio signal (L2 Speech) is fed into the Speech-to-Phonemes model (STP Model), the Accent Embedding and Gender Embedding model (AE/GE Model), and the Speaker Embedding model (SE Model). The accent vector (Accent Embedding) influences the generation of the phonetic representation, which is then fed into the Speech-to-Speech model (STS Model) to generate the mel-spectrogram. The mel-spectrogram generation is influenced by the same Accent Embedding, the Gender Embedding vector, and the individual vocal characteristics vector output by the SE Model. The generated spectrogram is then converted into the L1 Speech audio signal using the Vocoder Model.

The overall pipeline for generating L1 speech from the original L2 speech can be simplified into the following formula:

$$a_{L1} = F_V\left(F_{STS}\left(F_{STP}(a_{L2}, F_{AE}(a_{L2})), F_{AE}(a_{L2}), F_{GE}(a_{L2}), F_{SE}(a_{L2})\right)\right), \quad (1)$$

where $a_{L1}$ is the generated L1 speech audio signal; $a_{L2}$ is the input L2 speech audio signal; $F_V$ – Vocoder Model; $F_{STS}$ – STS Model; $F_{STP}$ – STP Model; $F_{AE}$ – AE/GE Model, Accent Embedding; $F_{GE}$ – AE/GE Model, Gender Embedding; $F_{SE}$ – SE Model, Speaker Embedding.

To create a unified end-to-end accent conversion model, each individual model must be trained sequentially. The AE/GE Model and SE Model are independent of other models and can be trained in any order. Training the STP Model requires the output of a pre-trained AE/GE Model. The STS Model requires all previous models (AE/GE, SE, STP) for its training. Finally, training the Vocoder Model requires the output of the STS Model.

### 2.1. Accent and Gender Embedding model (AE/GE Model)

To obtain fixed-length vectors representing the accent and gender properties of the speaker, the model was first trained to solve a classification task. In this configuration, class labels are used during training, which the model outputs at the last layer. The vector representations, used as vocal characteristics, are taken from a specific intermediate layer.

This and other models use the same Preprocessor. It is based on the Fast Fourier Transform (FFT), which converts the incoming signal from time domain into a frequency domain mel-spectrogram. This shows the frequency content of the audio signal over time on a perceptual mel scale, which approximates the non-linear frequency response of the human ear. The sampling rate is 22050 Hz, the window size is 1024 samples, the window hop size is 256 samples, and the number of generated mel bands is 80.

Figure 2 shows the training scheme of the AE/GE Model. It includes Jasper blocks configured as 3x3 [34]. Both the Accent Decoder and Gender Decoder have the same architecture, consisting of an Attention pooling layer [35], a normalization layer, a convolutional layer to obtain 192-dimensional vector representations (Accent Embedding, Gender Embedding), and a linear layer to predict the Accent Class or Gender Class.

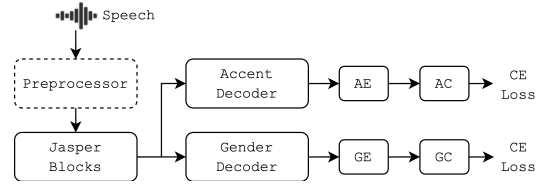

**Fig. 2.** Training scheme of the AE/GE Model.

The audio signal is converted into a mel-spectrogram, which then passes through Jasper blocks and, in parallel, through the Accent Decoder and Gender Decoder with corresponding fully connected layers at the output to obtain the accent and gender prediction vectors. During model training, the sum of cross-entropy losses is minimized:

$$L_{AE,GE}(x_a, y_a, x_g, y_g) = -\sum_{i=1}^{A} y_{a_i} \log\left(\frac{\exp(x_{a_i})}{\sum_{k=1}^{A} \exp(x_{a_k})}\right) - \sum_{j=1}^{G} y_{g_j} \log\left(\frac{\exp(x_{g_j})}{\sum_{l=1}^{G} \exp(x_{g_l})}\right), \quad (2)$$

where $L_{AE,GE}$ is the overall loss function of the AE/GE Model; $A$ – number of accent classes (40); $G$ – number of gender classes (2); $x_a$ – accent predictions; $x_g$ – gender predictions; $y_a$ – ground truth accent labels; $y_g$ – ground truth gender labels.

### 2.2. Speaker Embedding model (SE Model)

Figure 3 shows the training scheme of the SE Model. It includes an input convolutional neural network based on the SincNet architecture [36], layers from the X-Vectors DNN model [37], and a layer for obtaining 512-dimensional vector representations.

Unlike the AE/GE Model, no preliminary conversion to a mel-spectrogram is performed. Instead, the time-domain audio signal with a sampling rate of 16000 Hz is fed into the band-pass filters of the SincNet architecture, followed by the convo-lutional layers of the X-Vectors DNN, and the output fully connected layer. During model training the Additive Angular Margin (AAM) Loss [38] is minimized.

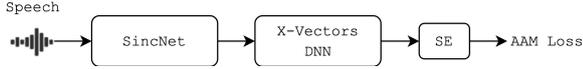

**Fig. 3.** Training scheme of the SE Model.

### 2.3. Speech-to-Phonemes model (STP Model)

The next step is speech recognition considering the speaker's accent. For this, a model that converts speech into phonetic or textual tokens is required. The training scheme of the STP Model is shown in Figure 4. In this figure, the blocks marked with dashed lines are fixed (or frozen) during backpropagation, meaning their weights are not updated.

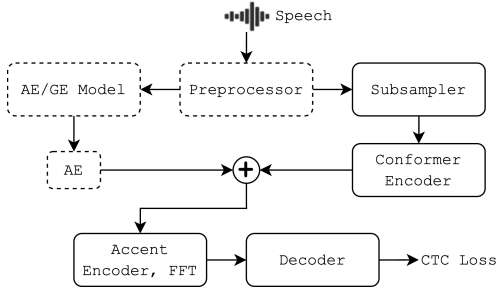

**Fig. 4.** Training scheme of the STP Model.

The incoming speech is fed into the Preprocessor, described earlier. Then, it is processed in parallel by the AE/GE Model to obtain the Accent Embedding (AE) and by the Subsampler block, which reduces the dimensionality by a factor of 4. It is then transformed in the Conformer Encoder block, which consists of 12 Conformer modules [39] with an internal dimensionality of 512, including fully connected, convolutional, and transformer layers. Next, the accent vector is normalized, adjusted to a dimensionality of 512, summed with the output of the Conformer Encoder, and fed into the Accent Encoder block. The Accent Encoder has a Feed-Forward Transformer (FFT) architecture [40]. The output from the Accent Encoder is used in the STS Model as the distribution of phonetic tokens. Finally, the output from the Accent Encoder is fed into a Decoder with a single-layer convolutional architecture and a Softmax activation function, producing a vector of predicted text tokens of a size equal to the tokenizer vocabulary (128) plus one (for the blank token). During model training, the Connectionist Temporal Classification (CTC) Loss function [41] is minimized, which calculates the loss between the continuous (unsegmented) time series and the target sequence:

$$L_{STP}(x,y) = -\log\left(\sum_{\rho \in A_{x,y}} \prod_{t=1}^{T} x_{\rho_t}\right), \quad (3)$$

where $L_{STP}$ is the loss function of the STP Model (CTC Loss); $x$ represents the probabilities of text tokens predicted by the model; $y$ – sequence of text tokens from the target text, $\rho$ – alignment path that reduces the $x$ predictions to the $y$ sequence by removing all blank tokens and merging repeating tokens; $A_{x,y}$ – set of all possible alignment paths; $T$ – number of predicted tokens in $x$; $x_{\rho_t}$ – probability of a specific predicted token at step $t$ for the selected alignment path $\rho$.

### 2.4. Spectrogram generation Speech-to-Speech model (STS Model)

The previous models are combined into a single architecture for speech-to-speech conversion and spectrogram generation. Figure 5 shows the training scheme of the STS Model. It includes the previously discussed Preprocessor, STP Model, as well as the AE/GE Model and SE Model blocks with their respective vector representation modules (AE, GE, SE). All these blocks are marked with dashed lines, indicating they were pre-trained and are not updated during the STS Model training. Additionally, a non-trainable block based on the Normalized Cross-Correlation Function and median smoothing is included for extracting the fundamental frequency (F0), which is the lowest frequency of a periodic sound signal perceived as pitch by the human ear [42].

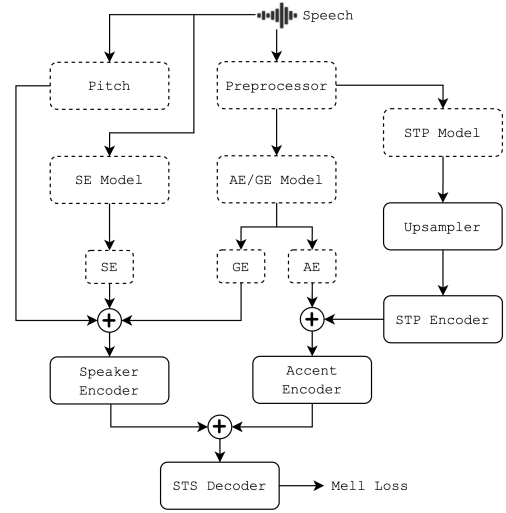

**Fig. 5.** Training scheme of the STS Model.

The incoming audio signal is processed by the Preprocessor, Pitch block, and SE Model. The mel-spectrogram from the Preprocessor is fed into the AE/GE Model and STP Model. The phonetic token distribution from the STP Model goes to the Upsampler block with a factor of 4 to align the original and generated spectrograms. The Upsampler consists of two 1D transposed convolutional layers and two ReLU activation functions, each placed after a convolutional layer. After the Upsampler, the phonetic representations are transformed by the STP Encoder, which has an architecture of six feed-forward Transformer (FFT) stacks [40] used in the Fastpitch architecture as an input block operating in the token domain [43], with internal and external dimensionalities of 1536 and 384, respectively.

The accent, gender, speaker timbre, and pitch profile vectors are normalized and adjusted to a dimensionality of 384. The accent vector and the output from the STP Encoder are summed and fed into the Accent Encoder (one FFT stack). Similarly, the pitch, timbre, and gender vectors are summed and fed into the Speaker Encoder (one FFT stack). Thus, the Speaker Encoder aggregates voice characteristics unrelated to the accent, while the Accent Encoder determines phonetic pronunciation based on the accent. The sum of the Speaker Encoder and Accent Encoder vectors is then fed into the STS Decoder, which has six FFT stacks of the Fastpitch architecture operating in the output mel-domain [43]. Finally, the vector is projected to a dimensionality of 80 to match the original number of mel bands. During training, the loss function based on the mean squared error is minimized:

$$L_{STS}(x,y) = \frac{1}{\sum_{i=1}^{N} d_i} \sum_{i=1}^{N} d_i (y_i - x_i)^2, \quad (4)$$

where $L_{STS}$ – STS Model loss function (Mel Loss); $N$ – number of elements (frames) in mel-spectrogram; $x$ – predicted mel-spectrogram; *y* – ground-truth target mel-spectrogram; *d* – mel-spectrogram duration mask, used to batch the data into a fixed size, consists of values 1 ("element should be considered") and 0 ("element should not be considered"), derived from the predicted spectrogram's duration.

### 2.5. Audio signal from spectrogram generation model (Vocoder Model)

The L1 speech mel-spectrogram in the frequency domain, obtained using the STS Model, is converted into an audio signal in the time domain. This is done using the HiFi-GAN model [44]. The output audio has a sampling rate of 22050 Hz. The model training is conducted as follows: audio from the training dataset is converted into a mel-spectrogram using the STS Model. The resulting mel-spectrogram is then passed to the Vocoder and converted into an audio signal. The loss functions for the generator and discriminator, as described in [44], are calculated using the generated and original audio.

### 2.6. Simplified accent conversion model (Ablation Model)

For the purpose of conducting comparative experiments, a simplified version of the accent conversion model was also developed. The scheme of this model is shown in Figure 6.

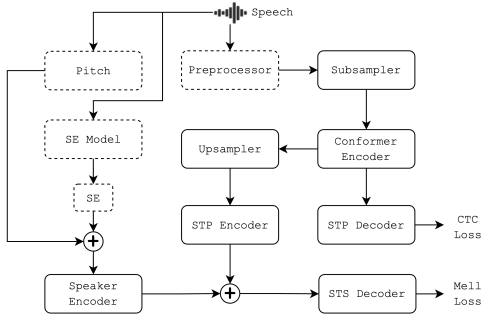

**Fig. 6.** Training scheme of the Ablation Model.

In this model, the vector representation module for accent and gender (AE/GE Model) and all associated encoders in the STP Model and STS Model are excluded. Thus, in the resulting Ablation Model, the output is not influenced by accent and gender properties. Additionally, the STP Model was not trained separately but simultaneously with the STS Model, without fixing the weights of the STP Model, minimizing the sum of the CTC Loss and Mel Loss functions.

### 3. EXPERIMENTAL SETUP

The AE/GE Model was trained on the following datasets: CMU-ARCTIC [45], L2 ARCTIC [46], Speech Accent Archive [47], Common Voice 16.1 [48]. These datasets consist of audio recordings of English speech, corresponding text transcripts, and additional metadata about accent, gender, and, in some cases, the speaker's native language, place of residence, and age. Using this information, the audio files were grouped into 40 classes representing native or foreign English accents, such as British, American, Russian, Indian and South Asian, Canadian, German, Australian, African, Japanese, Eastern European, etc. The gender of the speaker was also identified. The total duration of the labeled audio files was 1087.6 hours for the training set and 7.6 hours for the validation and test sets.

The SE Model was trained using the VoxCeleb1 [49] and VoxCeleb2 [50] datasets, with a total duration of 2794 hours. These datasets consist of grouped audio recordings of speech from 7363 individuals. Audio recordings from the same person are presented as positive examples during training, while recordings from different people are used as negative examples.

The STP Model was trained on the following datasets: CMU-ARCTIC [45], L2 ARCTIC [46], Common Voice 16.1 [48], LibriSpeech [51], NPTEL2020 [52], VCTK [53], and GigaSpeech [54]. These datasets consist of audio recordings of English speech with various accents and corresponding text transcripts. The total duration of the combined training set was 6107 hours, and the validation set was 48 hours. Text transcripts were normalized, i.e., converted from canonical written form to spoken form [55], which is especially important for numbers and abbreviations. They were also standardized: converted to lowercase, formatted in ASCII, with punctuation, special characters, and extra spaces removed. The training part of the texts was used to train the SentencePiece tokenizer [56] with a vocabulary size of 128, which was used to process all texts during model training and evaluation.

The STS Model and Vocoder were trained using the following datasets: CMU-ARCTIC [45], L2 ARCTIC [46], VCTK [53], LibriTTS-R [57], and LJ Speech [58]. The training and validation sets were divided with durations of 681 and 17.6 hours, respectively. During training, only audio information without text annotations was used.

The Ablation model was trained using the data for the STP Model and STS Model.

The code for training, evaluating, and using the described models was developed using the open-source libraries Pytorch [59] and NVIDIA NeMo [60]. The implementation and weights of the SE Model were taken from the Pyannote library [61]. Training was conducted on a server with eight NVIDIA Tesla V100 GPUs.

The AE/GE Model was trained using the SGD optimizer with a learning rate of $1*10^{-3}$, weight decay of $2*10^{-4}$, momentum of 0.9, and a Cosine Annealing scheduler for 200 epochs. The STP Model and STS Model were trained using the AdamW optimizer with a learning rate of $1*10^{-3}$, weight decay of 0.001, and the same scheduler as the AE/GE Model, for 50 epochs each. Fine-tuning of the HiFi-GAN vocoder model was performed using weights initialized from open sources [44], with the AdamW optimizer and a learning rate of $1*10^{-6}$ for 40 epochs. The Ablation model was trained with the same optimizer parameters, scheduler, and number of epochs as the STS Model.

**Table 1.** Number of trainable parameters

| Model | Million parameters |
|---|---|
| AE/GE | 24.9 |
| SE | 4.3 |
| STP | 82.1 |
| STS | 52.7 |
| **Full STS** | **164** |
| Vocoder | 84.7 |
| Total | 248.7 |

Table 1 shows the number of trainable model parameters. The entire accent conversion architecture (Full STS), consisting of several interconnected models, has a total of 164 million parameters.

### 4. RESULTS

#### 4.1. Performance evaluation

The model performance evaluation was conducted on a Linux server with one NVIDIA Tesla T4 GPU, eight vCPUs, and 16 GB of RAM. The model was first exported to the ONNX format and then deployed using the open-source NVIDIA Triton software. Using the API of the model deployed in NVIDIA Triton and a 5-second test audio file containing English L2 speech, latency measurements were performed over 200 iterations. The average generation latency was 52 milliseconds, and the throughput was 96 RTFX.

The performance evaluation results of the accent conversion model show low generation latency. Combined with the architectural feature that allows it to operate with segments shorter than 0.2 seconds without requiring long context accumulation, this makes the proposed model suitable for real-time dialogue applications where response delays affect communication [19, 20, 21].

### 4.2. Objective quality assessment

For the objective quality assessment, data from open sources and pre-trained speech recognition models were used. Using the proposed accent conversion method, an audio file was generated for each example in the test set. Quality metrics were then calculated for both the original and corrected audio files. The results of the objective quality assessment are presented in Table 2.

**Table 2.** Accent conversion model evaluation results using ASR Models

| Test dataset | ASR Model | | | |
|---|---|---|---|---|
| | Conformer | Citrinet | Whisper L. Mult. | Whisper M. En. |
| **WER, %** | | | | |
| ARCTIC | 9.57 | 11.73 | 16.23 | 8.91 |
| ARCTIC conv. | **8.78** | **11.55** | **12.69** | **8.68** |
| Common Voice | **9.07** | 25.80 | 36.89 | 11.26 |
| Common Voice conv. | 9.12 | **23.38** | **22.71** | **10.62** |
| NPTEL2020 | 29.18 | 29.88 | 16.41 | 15.18 |
| NPTEL2020 conv. | **25.26** | **29.41** | **13.87** | **11.64** |
| Afrispeech-200 | 43.2 | 46.24 | 37.91 | 33.61 |
| Afrispeech-200 conv. | **35.19** | **39.49** | **35.56** | **29.96** |
| **CER, %** | | | | |
| ARCTIC | 3.73 | 4.85 | 10.30 | 3.98 |
| ARCTIC conv. | **3.52** | **4.68** | **6.06** | **3.92** |
| Common Voice | **3.75** | 8.74 | 21.41 | 5.66 |
| Common Voice conv. | 3.77 | **8.29** | **13.63** | **5.22** |
| NPTEL2020 | 16.87 | 17.70 | 11.94 | 10.67 |
| NPTEL2020 conv. | **14.79** | **17.01** | **10.10** | **9.44** |
| Afrispeech-200 | 31.52 | 34.79 | 24.30 | 20.04 |
| Afrispeech-200 conv. | **27.86** | **28.92** | **23.15** | **18.88** |

The test datasets used for evaluation total 26.9 hours and were not included in the training process of the accent conversion model and its components. All of them consist of text transcripts and audio files of English speech with various native and non-native accents from open sources:

− 3.2 hours from CMU-ARCTIC [45], L2 ARCTIC [46] (ARCTIC), covering 10 accents: American, British, Chinese, Indian, Korean, Vietnamese, Spanish, Arabic, Dutch, German;

− 3.1 hours from Common Voice [48], covering 12 accents: American, British, Indian, Australian, African, Chinese, Filipino, Malaysian, German, Russian, French, Eastern European.

− 15.2 hours from NPTEL2020 [52], Indian accent;

− 5.4 hours from Afrispeech-200 [62], African accents: Yoruba, Swahili, Igbo, Zulu, Tswana, Idoma, Afrikaans.

Speech recognition models obtained from open sources were used: Conformer [39], Citrinet [63], Whisper [64]. The Whisper model was used in two variants: "large" multilingual (L. Mult.) and "medium" English-only (M. En.). Recognition was performed on both the unprocessed audio files and the audio files after accent conversion (conv.). The recognized and true transcripts were normalized [55], compared, and quality metrics were calculated: Word Error Rate (WER) and Character Error Rate (CER).

As seen from the results in the table, in almost all cases applying the accent conversion method improves the recognition of pre-trained models, as indicated by the lower error rates in words and characters. This means that the accent conversion model enhances speech quality, making it more recognizable.

### 4.3. Subjective quality assessment

Subjective listening tests were conducted with the participation of 53 people from different countries, each with an English proficiency level of at least B2 according to the CEFR scale. Each participant received instructions and was asked to listen to one or two audio files per experiment and rate the quality on a five-point scale, where "1" means "definitely does not meet the criterion," "2" means "probably does not meet," "3" means "compromise," "4" means "probably meets," and "5" means "definitely meets." The ratings were then used to calculate the Mean Opinion Score (MOS) for each experiment. The results are presented in Table 3.

**Table 3.** Subjective quality assessment results, MOS with 95% confidence interval

| Sample | Voice naturalness | Speaker similarity | Foreign accent absence |
|---|---|---|---|
| Original | 4.83 ± 0.10 | 4.91 ± 0.08 | 2.06 ± 0.18 |
| Ablation | 3.38 ± 0.13 | 3.92 ± 0.15 | 3.58 ± 0.17 |
| Proposed | 4.04 ± 0.16 | 4.30 ± 0.18 | 4.11 ± 0.14 |

Twenty pairs of audio files were randomly selected from the test subsets of the L2 ARCTIC [46] and NPTEL2020 [52] datasets with non-native English accents (Original): Indian, Chinese, Korean, Vietnamese, Spanish, Arabic, German. Each original audio pair consisted of recordings from the same speaker. For each selected audio file (a total of 40), variations were generated using the Ablation model (Ablation) and the Proposed model. Three experiments were conducted to assess voice naturalness, speaker similarity, and absence of foreign accent. In each experiment, participants were asked to make at least three ratings for each type of audio sample. The test samples themselves varied between experiments to avoid repetition. Participants could listen to each sample an unlimited number of times before giving a rating. Each respondent made a total of 9 to 12 ratings.

In the voice naturalness assessment, participants were asked to rate on a five-point scale how "natural" the speech in the audio sample sounded, i.e., whether the listener got the impression that it was a real human voice rather than synthesized or robotic speech. A score of "1" means the voice definitely sounds artificial and synthesized using computer generation methods, while a score of "5" means the speech sounds like it was recorded using analog or digital methods from a real human. Participants were also advised to ignore the presence or absence of background noise in the recording to focus solely on the speech quality.

For the speaker similarity assessment experiment, pairs of audio recordings were prepared: Original – Original, Original – Ablation, and Original – Proposed. The first pair consists of recordings only from the original data, representing the same speaker saying different phrases. The other pairs include an original recording of one phrase and a generated variant of another phrase by the same speaker. Participants were asked to listen to these pairs and determine if they were spoken by the same person, focusing on the timbre similarity between the files. A rating of "1" means the voices definitely belong to different people, while "5" means the timbre in the recordings is identical and belongs to the same person. Participants were instructed to ignore the accent properties (L1 or L2) to focus on comparing the voice timbre.

For the absence of a foreign accent assessment, participants were asked to listen to an English audio file and determine how pronounced the foreign accent was. British and American accents were considered native L1, while all others were non-native L2. A rating of "1" means the speech has a strong foreign L2 accent, while "5" means the speech is definitely native L1 English without a foreign accent.

The analysis of the table shows that the highest scores for voice naturalness and speaker similarity are given to the original examples, which is expected since they are not synthesized. This, along with the lowest score for the absence of a foreign accent, demonstrates the calibration of participants' opinions on real data. As indicated by the higher scores of the Proposed model compared to the Ablation model, the addition of the AE/GE Model significantly improves the generation quality. In all subjective experiments, the proposed model scores above "4," indicating that, according to the participants, the model likely meets the specified quality criteria.

## 5. CONCLUSIONS

A non-autoregressive real-time accent conversion model with voice cloning has been developed. This model consists of modules for accent and gender identification, speaker recognition, speech-to-phonetic token conversion, spectrogram generation, and decoding the resulting spectrogram into an audio signal. The model demonstrates high-quality accent conversion while preserving the original timbre and has low generation latency, making it suitable for real-time scenarios.

The proposed model allows to do the following:
1. Convert English speech with a foreign L2 accent to L1 speech without a foreign accent.
2. Improve speech quality, consequently enhancing the recognition quality of existing ASR systems.
3. Copy and change the speaker's vocal characteristics in real-time.
4. Convert L2 speech to L1 speech in real-time conversation mode.

The developed neural network model has demonstrated the capability to operate in English-language information systems in real-time. The research results can be applied in the development of voice modification systems as well as in speech recognition and speech generation systems.